\documentclass[preprint,12pt]{elsarticle}

\usepackage{amsmath}
\usepackage{amstext}
\usepackage{amssymb}
\usepackage{lineno,hyperref}
\usepackage{makecell}
\usepackage{color}
\usepackage{soul}

\usepackage{ulem}

\modulolinenumbers[5]

\biboptions{sort&compress}
\begin{document}

\begin{frontmatter}

\title{Constraining the invisible neutrino decay with KM3NeT-ORCA}


\author[IFIC,OK]{P.F.\ de Salas}
%\author{P.F.\ de Salas}
\ead{pabferde@ific.uv.es}
\author[IFIC]{S.\ Pastor}
%\author{S.\ Pastor}
\ead{pastor@ific.uv.es}
\author[IFIC]{C.A.\ Ternes}
%\author{C.A.\ Ternes}
\ead{chternes@ific.uv.es}
\author[IFIC]{T.\ Thakore}
%\author{T.\ Thakore}
\ead{tarak.thakore@ific.uv.es}
\author[IFIC]{M.\ T{\'o}rtola}
%\author{M.\ T{\'o}rtola}
\ead{mariam@ific.uv.es}  

\address[IFIC]{Institut de F\'{i}sica Corpuscular (CSIC-Universitat de Val\`{e}ncia), Parc Cientific de la UV\\
C/ Catedratico Jos\'e Beltr\'an, 2, E-46980 Paterna (Val\`{e}ncia), Spain}
\address[OK]{The Oskar Klein Centre for Cosmoparticle Physics,
Department of Physics, Stockholm University, SE-106 91 Stockholm, Sweden}

\begin{abstract}
Several theories of particle physics beyond the Standard Model consider that neutrinos can decay. 
In this work we assume that the standard mechanism of neutrino oscillations is altered by the decay of the heaviest neutrino mass state into a sterile neutrino and, depending on the model, a scalar or a Majoron. 
We study the sensitivity of the forthcoming KM3NeT-ORCA experiment to this scenario and find that it could improve the current bounds coming from oscillation experiments, where three-neutrino oscillations have been considered, by roughly two orders of magnitude. We also study how the presence of this neutrino decay can affect the determination of the atmospheric oscillation parameters $\sin^2\theta_{23}$ and $\Delta m_{31}^2$, as well as the sensitivity to the neutrino mass ordering.
\end{abstract}


\end{frontmatter}

%\linenumbers

\section{Introduction}
\label{sec:intro}

Over the last two decades or so, we have found overwhelming evidence for oscillating neutrinos. The oscillatory behavior can be described in terms of six parameters: the solar and atmospheric mass splittings $\Delta m_{21}^2$ and $\Delta m_{31}^2$, the solar angle $\theta_{12}$, the atmospheric angle $\theta_{23}$, the reactor angle $\theta_{13}$ and the CP phase $\delta$. These parameters have been measured in solar, reactor, atmospheric and long-baseline accelerator neutrino oscillation experiments. The level of precision reached by current experiments is such that, from the global picture~\cite{deSalas:2017kay,globalfit,Esteban:2016qun,nufit,Capozzi:2018ubv}, neutrino oscillation physics is entering the precision era. However, there are still some unknowns to be established:
\begin{enumerate}
 \item The true ordering of neutrino masses:  we still do not know if the order of the neutrino mass spectrum is normal (NO), where $m_3$ is the heaviest mass state, or inverted (IO), where $m_2$ is the heaviest one. 
Recent oscillation results show a preference for NO~\cite{deSalas:2017kay,globalfit}, although the real neutrino mass ordering is not fully determined yet~\cite{Gariazzo:2018pei}.
 \item The octant of the atmospheric angle: the measured value of $\sin^2\theta_{23}$ is close to maximal ($0.5$), but it can be either smaller (lower octant) or larger (upper octant).
 \item The exact value of the CP phase $\delta$: values of $\delta\approx 0.5\pi$ are now highly disfavored, but still a large part of the parameter space remains allowed. At 2$\sigma$ confidence level, CP might be maximally violated, but also conserved.
\item The absolute scale of neutrino masses: so far there are only upper bounds on it, coming from beta decay experiments and cosmological measurements~\cite{Aghanim:2018eyx}. 
\item The nature of neutrinos: are they Dirac or Majorana particles? In the latter case, there are two extra CP phases to be determined, which are only accessible through neutrinoless double beta decay experiments (see e.g.\ \cite{Vergados:2016hso}).
\end{enumerate}
The last two points can not be determined by  neutrino oscillation experiments, since flavor oscillations are insensitive to the absolute neutrino masses and to the Majorana CP phases.
Conversely, the three first issues are expected to be solved  by the future long-baseline experiment DUNE, which will measure very well the mass ordering~\cite{Acciarri:2015uup} as well as the value of the CP-violating phase and the octant of the atmospheric mixing angle~\cite{Srivastava:2018ser} within the standard three-neutrino picture. 

The determination of the mass ordering of neutrinos is also one of the main physics goals of the future atmospheric experiment ORCA (Oscillation Research with Cosmics in the Abyss)~\cite{Adrian-Martinez:2016fdl}, that will  provide precise measurements of the atmospheric parameters too.
ORCA will be basically an updated version of the ANTARES neutrino telescope, with a denser instrumented setup:  115 lines with 9 m spacing between the Digital Optical Modules (DOMs) and 20 m horizontal spacing between the detector lines. This will result in an energy threshold of only a few GeV, enhancing its sensitivity to lower energies with respect to ANTARES, that had a threshold of around 20 GeV. 
Therefore, ORCA seems a very promising candidate  not only to improve the current sensitivity to neutrino oscillation parameters, but also to look for signals of  physics beyond the Standard Model.
The presence of new physics  might change the well established  picture of neutrino oscillations and, hence,  it is crucial to improve  the precision of current measurements to look for potential deviations of the standard scenario, which would only arise at sub-leading order. 

One of these new physics scenarios, able to alter the neutrino oscillation pattern, is based on the existence of unstable neutrinos.
 Taking, for example, the Majoron model~\cite{Chikashige:1980ui,Gelmini:1980re,Schechter:1981cv,Gelmini:1983ea,GonzalezGarcia:1988rw},
   a neutrino $\nu_i$ can decay into a lighter neutrino $\nu_j$ and a new boson, the Majoron $J$, through $\nu_i\rightarrow\nu_j+J$ or $\nu_i\rightarrow\overline{\nu}_j+J$. 
 Likewise, Dirac neutrinos could decay through $\nu_{iL}\rightarrow\nu_{jR}+\xi$ into a scalar $\xi$ and a light right-handed neutrino $\nu_{jR}$. 
 If the decay product $\nu_j$ is an active neutrino, we talk about a \textit{visible} neutrino decay, otherwise it is an \textit{invisible}  decay.
 In this work we will focus  on the latter one. For previous studies on visible neutrino decay at current or future experiments, we refer the interested reader to Refs.~\cite{Eguchi:2003gg,Gago:2017zzy,Coloma:2017zpg,Ascencio-Sosa:2018lbk,Moss:2017pur,Pagliaroli:2015rca}.

Actually, the idea of unstable neutrinos is not new, since it was already proposed to explain the solar neutrino anomaly with a decaying mass state $\nu_2$~\cite{Bahcall:1972my}. 
However, neutrino decay alone could not explain the solar neutrino deficit and flavor oscillations were needed anyway~\cite{Acker:1993sz}. 
Therefore, it is usually assumed that this process can appear at subdominant level in combination with neutrino oscillations, as studied in Refs.~\cite{Berezhiani:1991vk,Berezhiani:1992xg,Bandyopadhyay:2001ct,Bandyopadhyay:2002qg}.
For the invisible neutrino decay, the best bound from oscillation experiments on the $\nu_2$ lifetime
comes from the combination of solar and reactor data and corresponds to approximately $\tau_2/m_2 > 2\times 10^{-3}$ s/eV at 90\% C.L.~\cite{Picoreti:2015ika}. See also Ref.~\cite{Berryman:2014qha} for a similar result and Ref.~\cite{Huang:2018nxj} for a recent forecast analysis on these parameters using dark matter detectors. 
 
The decay of the third neutrino mass eigenstate, $\nu_3$, was also considered as an attempt to explain the atmospheric neutrino problem, but it was found that also here it can contribute only with sub-leading effects~\cite{Lipari:1999vh}. Several studies using atmospheric and long-baseline experiments have been performed in this direction.
For instance, the analysis in Ref.~\cite{GonzalezGarcia:2008ru} combines neutrino data from Super-Kamiokande, K2K and MINOS to obtain the limit $\tau_3/m_3 > 2.9\times 10^{-10}$ s/eV at 90\% C.L.  
The authors of Ref.~\cite{Gomes:2014yua}, on the other hand, combine long-baseline neutrino data from T2K and MINOS,  obtaining the bound $\tau_3/m_3 > 2.8\times 10^{-12}$ s/eV at 90\% C.L.
Note, however, that these results have been derived under the two-neutrino approximation and, therefore,  a full three-neutrino analysis might loosen this bound. 
Recently, following a three-neutrino approach, a new constraint on the neutrino decay lifetime has been calculated from the combination of T2K and NO$\nu$A data ~\cite{Choubey:2018cfz}, giving approximately $\tau_3/m_3 > 2\times 10^{-12}$ s/eV at 90\% C.L. 

Prompted by the good sensitivity of the forthcoming experiment KM3NeT-ORCA to the atmospheric neutrino oscillation parameters in the GeV energy range, in this work we study
whether it can provide better bounds on the invisible neutrino decay, in comparison  to other neutrino oscillation experiments. We comment on current bounds from other astrophysical and cosmological probes in Sec.~\ref{sec:summary}.

This letter is organized as follows. Sec.~\ref{sec:invisible-decay} provides an introduction to the unstable neutrino scenario considered in this work. In Sec.~\ref{sec:ORCA}, we explain how the analysis and simulation of the ORCA experiment is performed. In Sec.~\ref{sec:discussion}, our main results are presented and discussed in detail. Finally, in Sec.~\ref{sec:summary}, we summarize the most relevant outcome of our work and give some concluding words. 

%%%%%%%%%%%%%%%%%%%%%%%%%%%%%%%%%%%%%%%%%%%%%%%%%%%%%%%%%%%%%%%%%%%%%%%%%
\section{Invisible neutrino decay}
\label{sec:invisible-decay}
\begin{figure}
\centering
\includegraphics[width=0.49\textwidth]{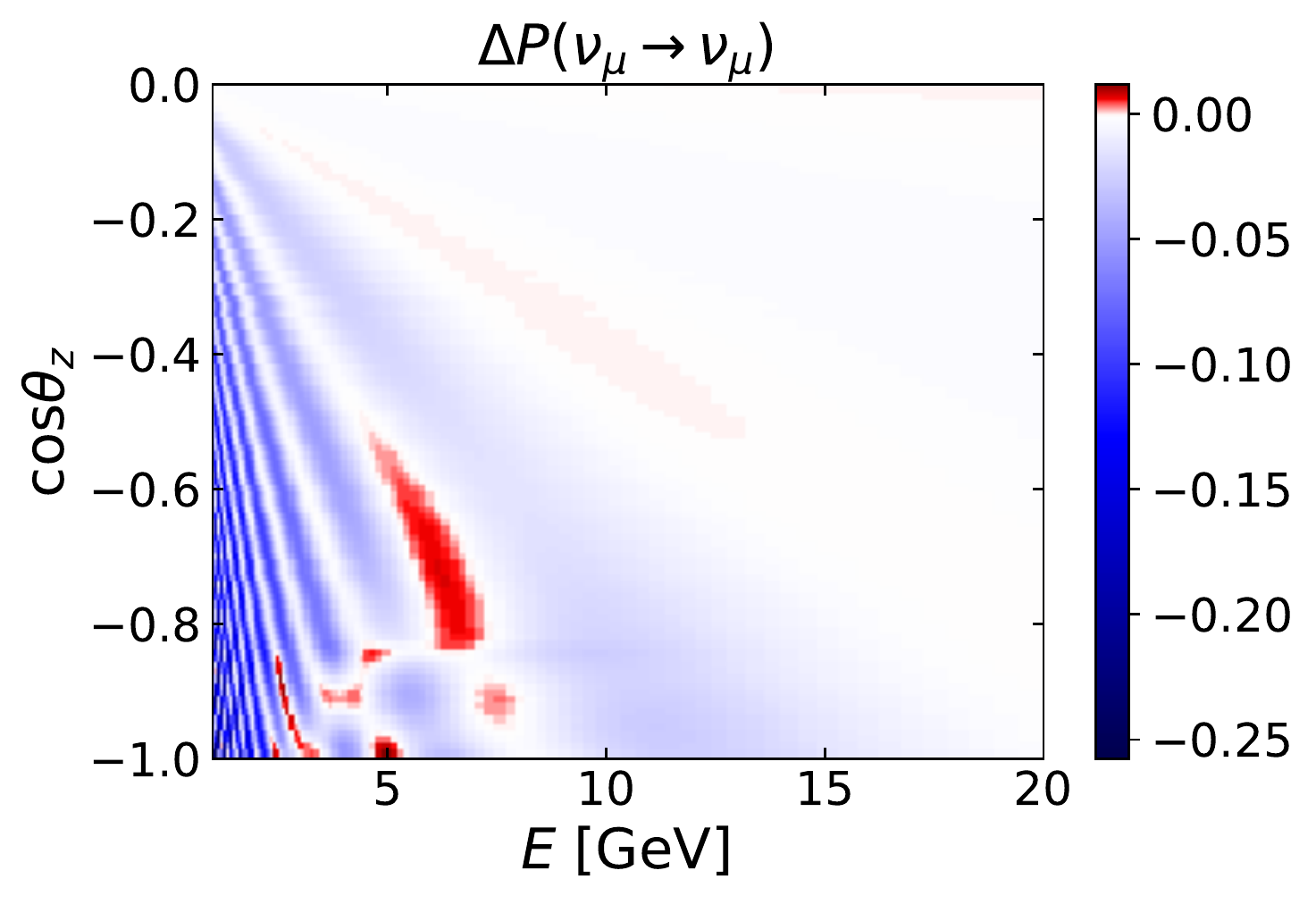}
\includegraphics[width=0.49\textwidth]{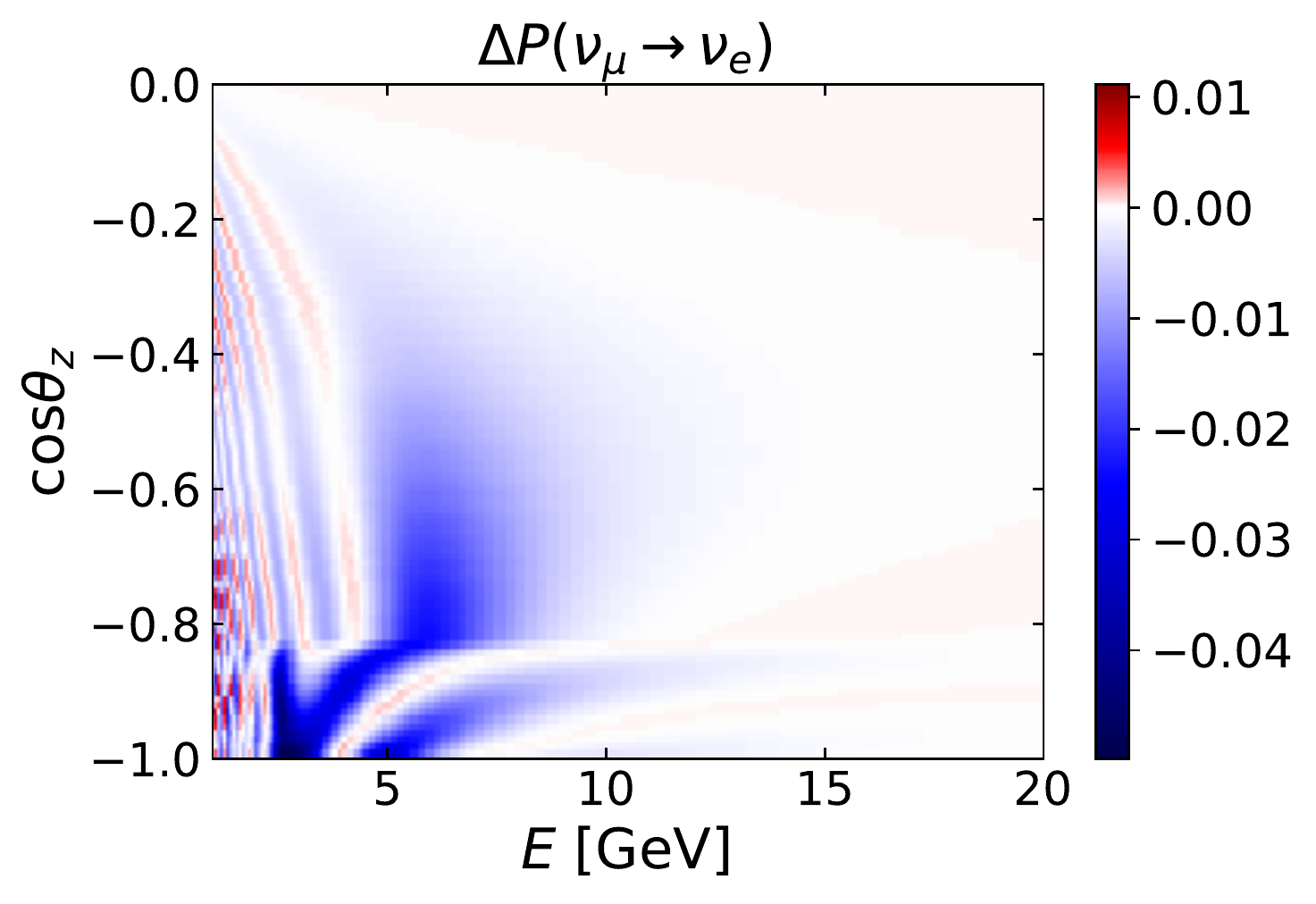}
\caption{\label{fig:oscil-alphas}
 Differences in the oscillation probability of atmospheric muon neutrinos with and without decay, $\Delta P = P^\mathrm{decay} - P^\mathrm{standard}$, for the $\nu_\mu \to \nu_\mu$ (left panel) and  $\nu_\mu \to \nu_e$ (right panel) oscillation channels. A  value of $\alpha_3 = 10^{-5}\,\mathrm{eV}^2$ has been assumed in both cases. The difference in the probabilities is shown as a function of the neutrino energy and  arrival zenith angle.}
\end{figure}

\begin{figure}
\centering
\includegraphics[width=0.7\textwidth]{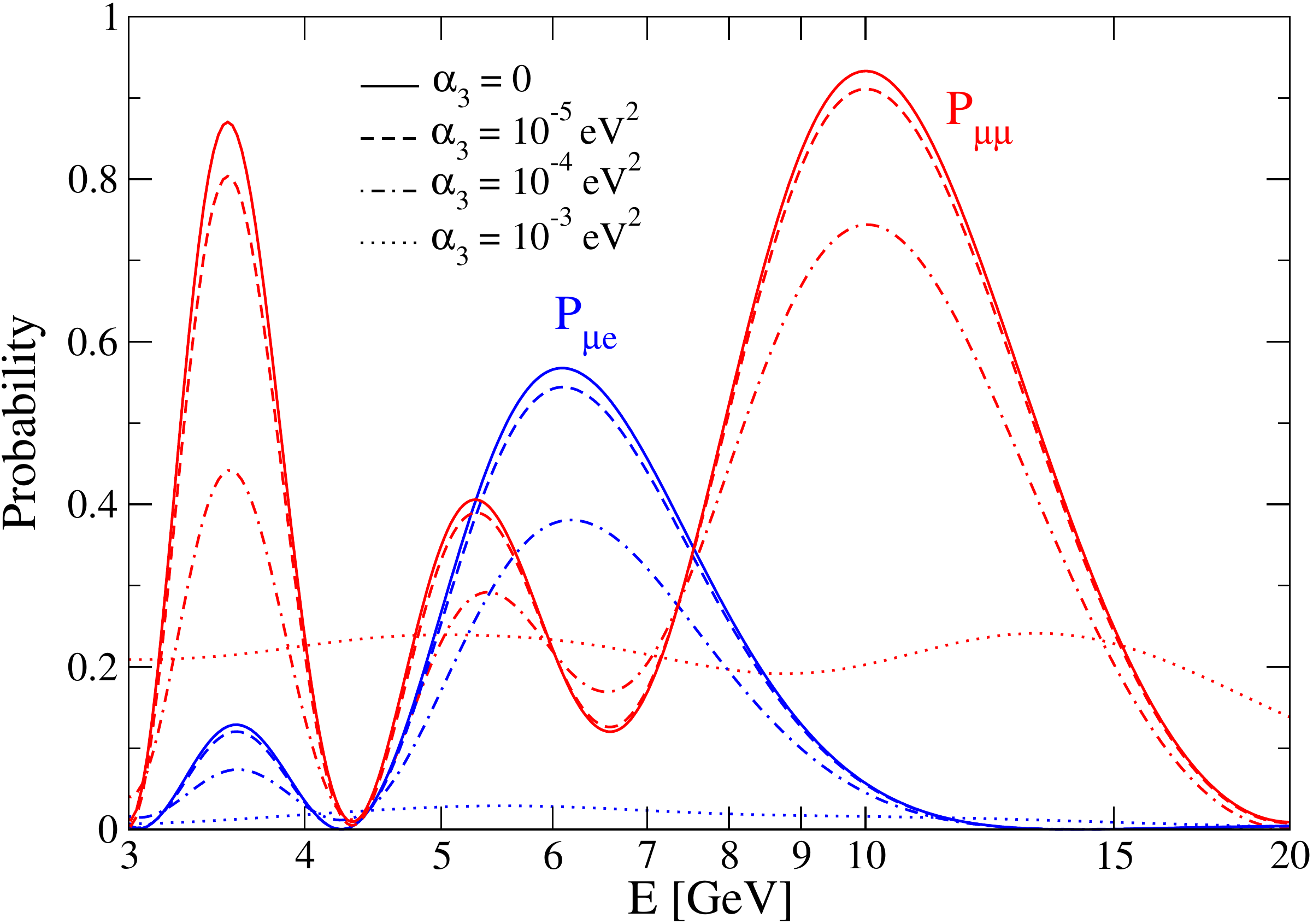}
\caption{\label{fig:oscil-prob} 
 Muon neutrino survival probability, $P_{\mu\mu}$, and electron neutrino appearance probability, $P_{\mu e}$, as a function of the energy for the arrival direction $\cos\theta_Z = -0.82$. The solid lines correspond to the standard case, without neutrino decay, while the dashed, dashed-dotted and dotted lines have been obtained assuming $\alpha_3 = 10^{-5}, 10^{-4}$ and $10^{-3}$ eV$^2$, respectively.
 }
\end{figure}

%%%%%%%%%%%%%%%%%%%%%%%%%%%%%%%%%%%%%%%%%%%%%%%%%%%%%%%%%%%%%%%%%%%%%%%%%
Here we discuss how the presence of an invisible neutrino decay would affect the calculation of the neutrino oscillation probability.
Besides the three light known neutrinos, we consider the presence of a fourth sterile neutrino, $\nu_4$.
Along this work, we will assume the  decay of the heaviest mass eigenstate ($\nu_3$ in NO) to this new neutrino state
\begin{equation}
 \nu_3\rightarrow \nu_4 + J,
\end{equation}
where $J$ is a pseudo-scalar singlet, or Majoron. 
We assume that there is no mixing among the three active neutrinos and the sterile one, so it can not oscillate back into an active state. Therefore,  the neutrino mixing matrix in vacuum is given by the standard three-family mixing matrix $U$,
\begin{equation}
 \begin{pmatrix}
  \nu_\alpha \\ \nu_s
 \end{pmatrix}
=
 \begin{pmatrix}
  U & 0 \\ 0 & 1
 \end{pmatrix}
\begin{pmatrix}
 \nu_k \\ \nu_4
\end{pmatrix},
\end{equation}
where the Greek index $\alpha = e,\mu,\tau$ indicates the flavor eigenstates and the Latin index $k=1,2,3$, the mass eigenstates. Because of the absence of active-sterile mixing, the propagation of the active states is not affected by the presence of $\nu_4$. For the mass spectrum we assume  normal mass ordering for the active states and a fourth state $\nu_4 = \nu_s$, lighter than the decaying one $m_4 < m_3$. Note that this is not in tension with the results of Refs.~\cite{Gariazzo:2017fdh,Gariazzo:2018mwd,Dentler:2018sju},  since we assume that the sterile state cannot oscillate back into active states. To take into account  the neutrino decay  in the  evolution process, we modify the neutrino Hamiltonian, including a decay constant $\alpha_3 = m_3 / \tau_3$, where $m_3$ is the heaviest neutrino mass and $\tau_3$ is its rest-frame lifetime. Hence, the full neutrino  Hamiltonian can be written as
\begin{equation}
 H = \frac{1}{2E} \left[H_0 + H_m + H_D\right],
 \label{Ham_decay}
\end{equation}
where the first two terms correspond to the standard  vacuum and  matter terms, namely
\begin{equation}
 H_0 = U
\begin{pmatrix}
 0&0&0
\\
 0&\Delta m_{21}^2 & 0
\\
 0&0&\Delta m_{31}^2
\end{pmatrix}
U^\dagger, \qquad
 H_m = 
\begin{pmatrix}
 V&0&0
\\
 0&0& 0
\\
 0&0&0
\end{pmatrix},
\end{equation}
with $V=2E\sqrt{2}G_F N_e$. $E$ is the neutrino energy, $G_F$ the Fermi constant and $N_e$ the electron number density. Finally, the last term 
in Eq.\ (\ref{Ham_decay}) represents the neutrino decay part
\begin{equation}
 H_D = U
\begin{pmatrix}
 0&0&0
\\
 0&0&0
\\
 0&0&-i\alpha_3
\end{pmatrix}U^\dagger .
\end{equation}
Then, effectively, the only change to the standard oscillation picture is a shift in the 33 entry of the Hamiltonian in the mass basis, from $\Delta m_{31}^2$ to $\Delta m_{31}^2 -i\alpha_3$. 
Note, however,  that this implies that the sum of the neutrino oscillation probabilities might be different from one,
\begin{equation}\label{eq:Psum-decay}
 P_{\alpha e} + P_{\alpha \mu} +P_{\alpha \tau} < 1,\quad \alpha=e,\mu,\tau.
\end{equation}
Therefore, if the heaviest neutrino mass eigenstate $\nu_3$ decays, apart from a changed oscillatory pattern, we could also have missing neutrinos.
In order to show the effect of the decay, we present in Fig.~\ref{fig:oscil-alphas} the difference in the survival probability (left panel) of atmospheric muon neutrinos with and without decay, $\Delta P_{\alpha\beta} = P_{\alpha\beta}^\mathrm{decay} - P_{\alpha\beta}^\mathrm{standard}$, for a value of $\alpha_3 = 10^{-5}\,\mathrm{eV}^2$.  The right panel shows the analogous result for the electron neutrino appearance probability.
 Note that this value of the decay constant is rather large and we have chosen it for illustrative purposes only. 
 As it is seen in the figures, the main effect of the decay is concentrated in the region of neutrino energies close to the resonance ($\sim$ 3--8$\,\mathrm{GeV}$) for values of the zenith angle at which matter effects are more relevant. This results in a softening of the oscillation pattern, as better illustrated in Fig.~\ref{fig:oscil-prob}. There, we  show the muon neutrino survival probability $P_{\mu\mu}$ as well as the electron neutrino appearance probability $P_{\mu e}$ for a particular arrival direction, corresponding to $\cos\theta_Z = -0.82$. 
 The solid lines in the plot correspond  to the standard case, without neutrino decay, while the dashed, dashed-dotted and dotted lines have been obtained assuming values of the decay constant $\alpha_3$ equal to  $10^{-5}$, $10^{-4}$ and $10^{-3}$ eV$^2$, respectively.
 In this figure, one can see how the presence of the invisible neutrino decay leads to a reduction of the oscillatory behavior, that becomes almost suppressed for larger values of $\alpha_3$, with special impact for the values of $\cos\theta_Z$ and neutrino energies close to the matter effect resonance at around 6--7 GeV. 
 %
% The resonance due to the matter effects is visible for neutrino energies around 6--7 GeV.
  An interesting result from this suppression is the increment of the survival probability at some regions in the plane $\cos\theta_Z$--$E$, which seems contradictory with the idea of decaying neutrinos. This comes from the softening of flavor oscillations in the presence of the $\nu_3$ decay, since it washes out oscillations controlled by $\Delta m_{31}^2$ while keeping untouched those driven by $\Delta m_{21}^2$. As a consequence, an almost averaged oscillation pattern appears, which enhances the survival muon neutrino probability at its minima. However, Eq.~(\ref{eq:Psum-decay}) holds and the sum of neutrino probabilities below unity tells us that they are actually decaying.

%%%%%%%%%%%%%%%%%%%%%%%%%%%%%%%%%%%%%%%%%%%%%%%%%%%%%%%%%%%%%%%%%%%%%%%%%
\section{Numerical analysis}
\label{sec:ORCA}
%%%%%%%%%%%%%%%%%%%%%%%%%%%%%%%%%%%%%%%%%%%%%%%%%%%%%%%%%%%%%%%%%%%%%%%%%

In this section, we present the numerical procedure followed to simulate the neutrino signal in  ORCA. 
First, we explain how to calculate the number of events and the $\chi^2$ functions. Next, we provide some details about how to handle the systematic uncertainties in the data analysis.

\subsection{Simulation of the neutrino signal in ORCA}

The simulation of the neutrino signal expected in ORCA requires the knowledge of the atmospheric neutrino flux, that we take from \cite{Honda:2015fha} and
 further modulate with flavor oscillations. 
The conversion probability,  from the neutrino creation point in the atmosphere to the detector after traversing the Earth, is numerically calculated considering three-neutrino oscillations in matter. 
In order to do so, we discretize the neutrino path and consider the matter density at each point using the PREM profile~\cite{Dziewonski:1981xy}.
 
The sensitivity tests presented in this work are performed  simulating the number of events detected by ORCA in a binned area of the  
parameters $\cos\theta_{z,{\rm rec}}$ and $\log_{10}\left(E_{\nu,{\rm rec}} / \mathrm{GeV} \right)$, where $\theta_{z,{\rm rec}}$ is the reconstructed zenith angle\footnote{The zenith angle is defined such that $\theta_z =0$ corresponds to vertical down-going events.} 
and $E_{\nu,{\rm rec}}$ is the reconstructed neutrino energy. 
Following the indications in Ref.~\cite{Adrian-Martinez:2016fdl}, we divide the reconstructed parameter ranges, $\cos\theta_{z,{\rm rec}}\in [-1 , 0]$ and $\varepsilon_{\rm rec} = \log_{10}\left(E_{\nu,{\rm rec}} / \mathrm{GeV} \right)$, in 20 bins each, where $E_{\nu,{\rm rec}} \in [1,21]\,\mathrm{GeV}$. 
We have also considered larger values for the maximum reconstructed neutrino energy, but the results are essentially unchanged.

Given the incapability of ORCA to distinguish neutrinos from antineutrinos, both contributions are summed in each bin. Our analysis, however, makes a distinction 
between the two different topologies produced in a neutrino interaction with the sea water molecules, namely \textit{track-like} or 
\textit{shower-like}\footnote{Interestingly, invisible neutrino decay has been recently suggested as a way to explain the tension between the event topologies (tracks or cascades) of the detected high-energy astrophysical neutrinos at the IceCube observatory \cite{Denton:2018aml}.}. A track is an elongated signal of deposited energy in ORCA's photomultipliers (PMTs), which is mostly produced when a $\nu_\mu$ (or its antiparticle) interacts through charged-current (CC) interactions, producing a muon that travels a long distance before losing all its energy. The same topology can be produced in a $\nu_\tau$ CC interaction if the generated tau decays into a muon, which produces the track. Therefore, this constitutes an unavoidable background for 
muon neutrinos detected through CC. On the other hand, in all other cases ($\nu_e$ CC, $\nu_\tau$ CC with the tau not decaying into a muon and all flavor neutral-current interactions) the neutrino energy is quickly lost into an electromagnetic cascade, a hadronic cascade or both, depending on the interaction, giving rise to a shower-like topology. In current large-volume neutrino telescopes, like ANTARES or IceCube, the directionality of an event producing a shower is measured with a large uncertainty. However, thanks to the multi-PMT cha\-rac\-teristic of ORCA's DOMs, the directionality of a shower-like event will be known with a precision of a few degrees, where the exact accuracy depends on the incoming neutrino energy (less than 10 degrees for energies larger than $5\,\mathrm{GeV}$) \cite{Adrian-Martinez:2016fdl}.

The expected number of events per bin, $N^{c\alpha }_{ij}$, corresponding to a given interaction channel, $c$, and a neutrino flavor, $\alpha$, contributing to a given topology, is calculated from the convolution of the neutrino flux at the detector with the corresponding cross section, detector resolutions and detector effective mass,
\begin{align}\label{eq:Nev-alpha}
N^{c\alpha }_{ij} =& \, 2\pi t \ln^2 (10) 
\int^{\cos\theta_{z,\mathrm{rec}}^{i+1}}_{\cos\theta_{z,\mathrm{rec}}^{i}} \mathrm{d}x 
\int^{\varepsilon_{\mathrm{rec}}^{j+1}}_{\varepsilon_{\mathrm{rec}}^{j}} \mathrm{d} y 
\int^{1}_{-1} \mathrm{d}\cos\theta_{z} \, \nonumber\\
\times & \int_{-\infty}^{+\infty} \mathrm{d}\varepsilon_{\rm true} \, E_{\rm rec}(y)E_{\rm true} (\varepsilon_{\rm true})
\frac{M^{c\alpha}_{\rm eff} \left( \varepsilon_{\rm true} \right)}{m_p}  \, \nonumber\\
\times &
R^{c\alpha} \left( x,y, \cos\theta_{z}, \varepsilon_{\rm true}\right) 
\frac{\mathrm{d}^2 \phi^{c\alpha}_{\rm det}}{\mathrm{d}\cos\theta_{z} \mathrm{d}E_{\rm true}}\left( \cos\theta_{z}, \varepsilon_{\rm true} \right),
\end{align}
where $t$ is the total time of data acquisition, $m_p$ the proton mass, $\varepsilon_{\rm true} = \log_{10}\left(E_{\nu,{\rm true}}/\mathrm{GeV}\right)$. $M^{c\alpha}_{\rm eff}$, $R^{c\alpha}$ and $\phi^{c\alpha}_{\rm det}$ are the detector effective mass, the detector resolution and the number of neutrinos per second at the detector for the corresponding interaction channel and neutrino flavor, respectively.
The indices $i$ and $j$ refer to the $i$th bin in reconstructed zenith angle $\theta_{\rm z,rec}$ and the $j$th bin in reconstructed energy $\varepsilon_{\rm rec}$. The number of $\nu_\alpha$ per second at the detector for the interaction channel $c$ is given by
\begin{equation}
\frac{\mathrm{d}^2 \phi^{c\alpha}_{\rm det}}{\mathrm{d}\cos\theta_{z} \mathrm{d}E_{\rm true}} = 
\sigma^{c\alpha} \sum_{\beta =\{e,\mu\}} \frac{\mathrm{d}^2\phi_{\beta}^{0}}{\mathrm{d}\cos\theta_{z} \mathrm{d}E_{\rm true}} P_{\nu_\beta \rightarrow \nu_\alpha},
\end{equation}
where $\sigma^{c\alpha}$ is the cross section for $\nu_\alpha$ in the interaction channel $c$~\cite{Messier:1999kj,Paschos:2001np}, $\phi_{\beta}^0$ is the atmospheric neutrino flux~\cite{Honda:2015fha} and $P_{\nu_\beta \rightarrow \nu_\alpha}$ is the probability for a $\nu_\beta$ to oscillate into a $\nu_\alpha$ when arriving at ORCA.
To simulate the atmospheric neutrino signal in ORCA, we have fixed the values of the neutrino oscillation parameters to their best-fit values found in 
\cite{deSalas:2017kay,globalfit} and summarized in Tab.~\ref{tab:datasim}. 
Note that, to establish ORCA's sensitivity to the invisible neutrino decay, we  set the decay parameter $\alpha_3$ to zero in the simulated data. 

\begin{table}[t!]\centering
  \catcode`?=\active \def?{\hphantom{0}}
   \begin{tabular}{|c|c|}
    \hline
    parameter & value
    \\
    \hline
    $\Delta m^2_{21}$& $7.55\times 10^{-5}$ eV$^2$\\  
    $\Delta m^2_{31}$&  $2.50\times 10^{-3}$ eV$^2$\\
    $\sin^2\theta_{12}$ & 0.32\\ 
     $\sin^2\theta_{23}$ & 0.547\\
    $\sin^2\theta_{13}$ & 0.0216\\
   $\delta$ & 1.32$\pi$\\
   \hline
   $\alpha_3$ & 0\\       
   \hline
     \end{tabular}
     \caption{ \label{tab:datasim} 
        Neutrino oscillation parameters and decay constant used to simulate the atmos\-pheric  data in ORCA.}
\end{table}

The detector resolutions in Eq.~\ref{eq:Nev-alpha}, $R^{c\alpha}$,  include both the zenith angle and energy resolutions
\begin{equation}
R^{c\alpha}\left(\theta_{z,{\rm rec}} , E_{\nu,{\rm rec}} ,\theta_{z,{\rm true}} , E_{\nu,{\rm true}} \right) 
= r^{c\alpha}_{E_\nu} (E_{\nu,\mathrm{rec}}, E_{\nu,{\rm true}})\; 
r^{c\alpha}_{\theta_z} (\theta_{z,\mathrm{rec}},\theta_{z,{\rm true}},E_{\nu,{\rm true}}),
\label{eq:det-response}
\end{equation}
where the dimensions of the individual resolutions are given by $\left[ r^{c\alpha}_{E_\nu} \right] = \mathrm{GeV}^{-1}$ and $\left[ r^{c\alpha}_{\theta_z} \right] = \mathrm{rad}^{-1}$. For simplicity, here we have neglected the dependence of the energy resolution on the arrival direction, that is actually very small~\cite{Adrian-Martinez:2016fdl}.

Finally, with all these ingredients, one can calculate the total number of events for a given topology (track-like or shower-like) expected in ORCA. The total event number is obtained  by convolving each interaction channel with the corresponding particle identification performance: $T^{c\alpha}_{\rm pid}$ for tracks and $S^{c\alpha}_{\rm pid}$ for showers. Each of them represents the probability that ORCA identifies an event produced from a $\nu_\alpha$ via an interaction channel $c$ with the given topology. Thus, the number of events inside the $ij$-bin identified with a topology $\mathcal{T}$  is
\begin{equation}
N_{ij}^{\mathcal{T}} = \sum_{\alpha=\{e,\mu,\tau\}} \sum_c \mathcal{T}^{c\alpha}_{\rm pid} N_{ij}^{c\alpha},
\end{equation}
with $\mathcal{T}^{c\alpha}_{\rm pid} = T^{c\alpha}_{\rm pid},S^{c\alpha}_{\rm pid}$.
The detector-dependent quantities ($M^{c\alpha}_{\rm eff}$, $R^{c\alpha}$ and $\mathcal{T}^{c\alpha}_{\rm pid}$) corresponding to ORCA have been obtained by fitting the information provided in \cite{Adrian-Martinez:2016fdl} for a configuration of 9 m spacing between DOMs in a line, chosen as the final experimental setup  by the KM3NeT collaboration.

%%%%%%%%%%%%%%%%%%%
%%%%%%%%%%%%%%%%%%%
\subsection{The role of systematics}
\label{sec:analysis}

We estimate ORCA's sensitivity to the invisible neutrino decay defining a $\chi^2$ function in terms of the most relevant parameters: the decay constant $\alpha_3$ and the atmospheric parameters, $\Delta m_{31}^2$ and $\sin^2\theta_{23}$. 
With this $\chi^2$, we fit the simulated signal in ORCA, obtained as explained in the previous subsection. 
Besides the details commented there, one should also consider the presence of systematic uncertainties that may affect the simulation of the experiment. 
For instance, the uncertainties on the determination of the atmospheric neutrino flux or the limited knowledge of the detector response will certainly modify the calculation of the expected number of events in a given experiment.
These systematic uncertainties are usually included  in the numerical analysis with some nuisance parameters $\epsilon_i$, implemented in the definition of the $\chi^2$ function
\begin{align}
 \chi^2 (\sin^2\theta_{23},\Delta m_{31}^2,\alpha_3) = & \min_{\vec{\epsilon}}\left\{\sum_{i,j} \left(\frac{N_{ij}(\sin^2\theta_{23},\Delta m_{31}^2,\alpha_3;\vec{\epsilon}) - N_{ij}^\text{dat}}{\sqrt{N_{ij}^\text{dat}}}\right)^2 \right. \nonumber\\
 +& \left.\sum_k \left(\frac{\epsilon_k - \mu_k}{\sigma_k}\right)^2 \right\}.
 \label{eq:chi2}
\end{align}
Here $i$ ($j$) indicates the $i$th ($j$th) bin in azimuth angle (energy), $N_{ij}^\text{dat}$ is the simulated number of events in this bin (in analogy with the observed one for a running experiment) and $N_{ij}(\sin^2\theta_{23},\Delta m_{31}^2,\alpha_3;\vec{\epsilon})$ is the event number for the oscillation parameters $\sin^2\theta_{23}$ and $\Delta m_{31}^2$, the decay constant $\alpha_3$ and the nuisance parameters $\vec{\epsilon}=(\epsilon_1,\epsilon_2,\ldots)$.
These  parameters are fitted after the minimization of the $\chi^2$ function, that includes a pull factor  penalizing large deviations from their corresponding expectation values, $\mu_k$, compared to the associated errors, $\sigma_k$.
\begin{table}[tb!]
\centering
\begin{tabular}{|c|c|c|}
\hline
Systematic & \makecell{Expectation\\value ($\mu_k$)} & \makecell{Standard\\deviation ($\sigma_k$)} \\
\hline
$\mathcal{N}$ & $1$ & flat\\
$\gamma$ & $0$ & flat \\
$f_{\mu e}$ & $0$ & $0.1$ \\
$f_{\nu / \bar{\nu}}$ & $0$ & $0.1$\\
\hline
$f_{r,{\rm shower}}$ & $0$ & $0.2$ \\
$f_{r,{\rm track}}$ & $0$ & $0.2$ \\
$f_{E}$ & $0$ & $0.01$
\\\hline
\end{tabular}
\caption{List of systematics used to reproduce ORCA's functionality and uncertainties in the atmospheric neutrino fluxes, where $\mu_k$ and $\sigma_k$ are the corresponding central value and gaussian dispersion, respectively.
}
\label{tab:systematics}
\end{table}
The systematics used in our analysis are listed in Tab.~\ref{tab:systematics}, together with their corresponding expectation values and errors.
As explained above, we consider systematic uncertainties related to the  detector functionality and to the theoretical predictions of the  atmospheric neutrino fluxes. 
The first four entries in Tab.~\ref{tab:systematics} are related to the atmospheric neutrino flux as, for instance, a global normalization uncertainty, $\mathcal{N}$.  We also consider $\gamma$, the spectral index (with a pivot point in neutrino energy at $20\,\mathrm{GeV}$), and two systematics related to the composition of the atmospheric flux, namely the fraction of muon neutrinos to electron neutrinos $\phi_{\nu_\mu}/\phi_{\nu_e}$ in the original fluxes: $f_{\mu e}$, and another one regarding the fraction of neutrinos to antineutrinos $\phi_{\nu}/\phi_{\bar{\nu}}$: $f_{\nu/\bar{\nu}}$. These last two modify the flux composition, such that $\phi\rightarrow\tilde{\phi}$. In combination with the other systematics, the flux is changed as
\begin{equation}
 \phi(E,\theta_z)\rightarrow \mathcal{N}\tilde{\phi}(E,\theta_z)\left(\frac{E}{E_{\text{pivot}}}\right)^\gamma.
\end{equation}
The rest of the systematics ($f_{r,{\rm shower}}$, $f_{r,{\rm track}}$ and $f_{E}$) are associated to detector-related effects that might affect, respectively, the resolution of shower and track events, as well as the reconstructed energy. The first two simply modify the detector response function in Eq.~(\ref{eq:det-response}), discussed in the last subsection, while $f_E$ modifies the incoming (or true) neutrino energy.  This sys\-te\-matic uncertainty affects the event number calculation in a more complicated way,  since it replaces $E_{\rm true}$ by $E_{\rm true}(1+f_E)$ everywhere in the simulation and, therefore, modifies directly the oscillation probabilities. 
Note that all the systematics named with an $f$ account for small deviations from their corresponding central value. Therefore, their effect can be accounted for by modifying the corresponding variable $X$ to which they affect, such that $X \rightarrow X (1 + f_X )$.

%%%%%%%%%%%%%%%%%%%%%%%%%%%%%%%%%%%%%%%%%%%%%%%%%%%%%%%%%%%%%%%%%%%%%%%%%
\section{Results and discussion}
\label{sec:discussion}
%%%%%%%%%%%%%%%%%%%%%%%%%%%%%%%%%%%%%%%%%%%%%%%%%%%%%%%%%%%%%%%%%%%%%%%%%

In this section, we present the results of our analysis, assuming that the ORCA experiment is running for three or ten years. First, we present the bounds it could put on the invisible decay of $\nu_3$ from the observation of atmospheric neutrino oscillations. Next, we show how the invisible neutrino decay can affect the determination of the atmospheric neutrino parameters when marginalizing over $\alpha_3$. The systematic errors considered in this analysis were summarized in Sec.~\ref{sec:analysis}. Since current global data prefer normal neutrino mass ordering~\cite{Gariazzo:2018pei,deSalas:2018bym}, here we will focus only on this case.

\subsection{Estimated sensitivity to the invisible decay}

\begin{figure}
\centering
\includegraphics[width=0.85\textwidth]{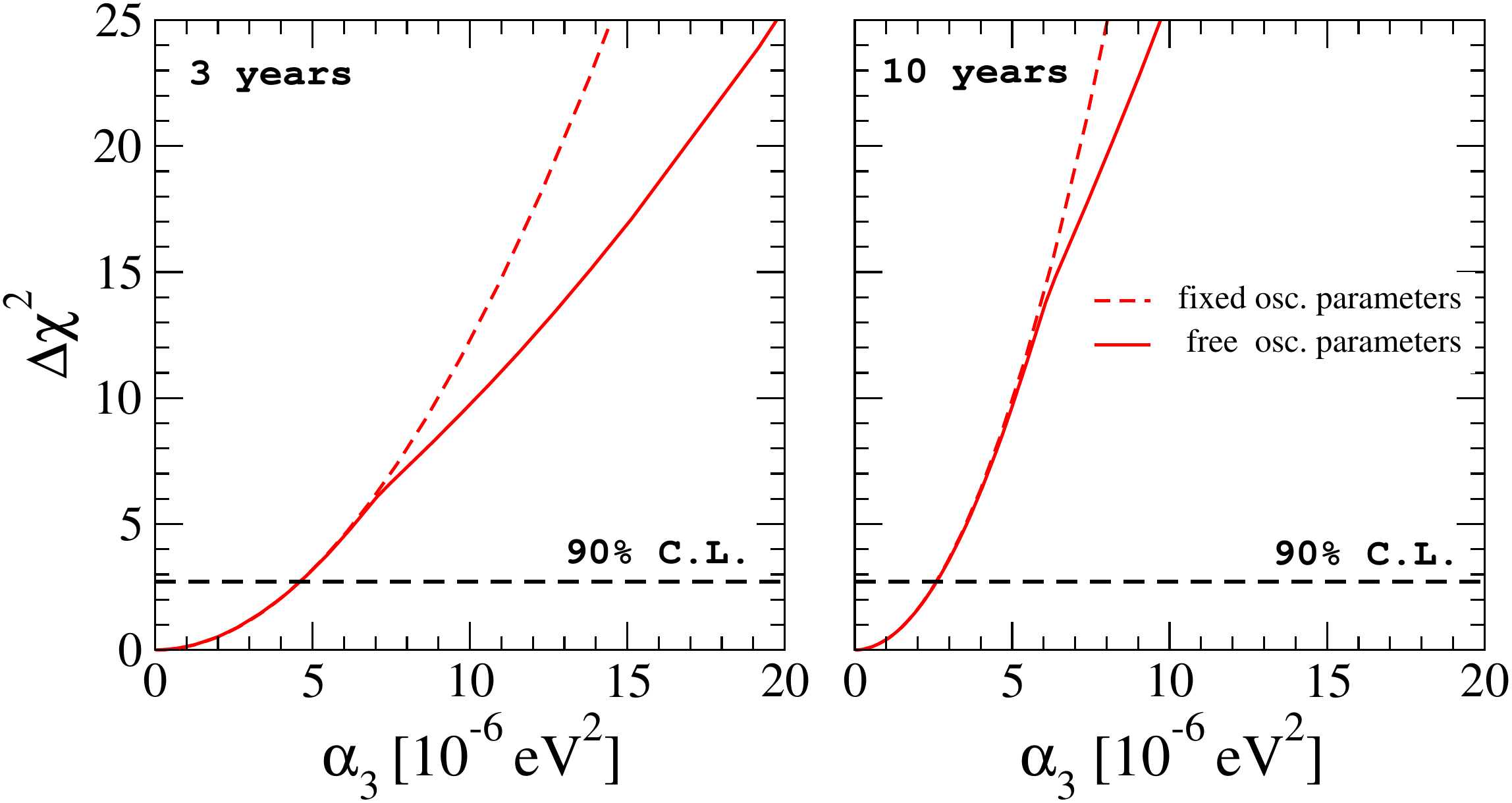}
\caption{
Sensitivity of ORCA to  the neutrino decay constant $\alpha_3$. Solid (dashed) lines correspond to the analysis with free (fixed)  oscillation parameters $\Delta m_{31}^2$ and $\theta_{23}$. 
The left (right) panel shows the results for three (ten) years of operational time.}
\label{fig:alpha3sensitivity}
\end{figure}

Our results for the sensitivity to the decay constant $\alpha_3$ are presented in Fig.~\ref{fig:alpha3sensitivity}. 
We have followed two approaches to estimate ORCA's sensitivity to the invisible decay.  First, we have fixed the oscillation parameters, $\Delta m_{31}^2$ and $\sin^2\theta_{23}$, to their best fit point and we have calculated the $\chi^2$ function following Eq.~(\ref{eq:chi2}). The results of this analysis are reported with dashed lines in Fig.~\ref{fig:alpha3sensitivity}, where we have 
considered 3 and 10 years of running time. Then, we have followed a more realistic procedure, 
minimizing the $\chi^2$ function defined in Eq.(\ref{eq:chi2}) over the oscillation parameters for each value of $\alpha_3$,
\begin{equation}
 \chi^2_{\text{decay}}(\alpha_3) = \min_{\sin^2\theta_{23},\Delta m_{31}^2}\left[\chi^2 (\sin^2\theta_{23},\Delta m_{31}^2,\alpha_3)\right]\, .
\end{equation}
The minimization is performed over all possible values of $\Delta m_{31}^2$ and $\sin^2\theta_{23}$, where no priors on the parameters have been used.
Our findings for this case correspond to the solid lines plotted in Fig.~\ref{fig:alpha3sensitivity}.
Thanks to the expected good resolution of ORCA to $\Delta m^2_{31}$ and $\theta_{23}$, the prior knowledge on the oscillation parameters  affects   
 the sensitivity to neutrino decay only for very large values of $\alpha_3$, as can be seen by comparing the solid and dashed lines in the two panels of Fig.~\ref{fig:alpha3sensitivity}. This worsening of the sensitivity comes mostly from a jump in the preferred  value of $\theta_{23}$ from the second octant (corresponding to the best-fit value of $\theta_{23}$, used to simulate ORCA's data)  to the first one.  
We have tested that the change of behavior in the sensitivity curves actually disappears when a value of $\sin^2\theta_{23} <  0.5$ is chosen to simulate the fake data. The reason of this feature is somewhat subtle, but it is related to the fact that larger modifications of the oscillation probabilities due to neutrino decay are expected close to the matter effect resonance, as explained in Sec.~\ref{sec:invisible-decay}. 
In this region, there is a large influence coming from the neutrino appearance probabilities, $P_{\nu_{e, \mu} \rightarrow \nu_{\mu ,e}}$,   mainly driven by a term proportional to $\sin^2\theta_{23}$ (contrary to the muon neutrino disappearance probability,  dominated by $\sin^2 2\theta_{23}$)~\cite{Adrian-Martinez:2016fdl}. 
Because a larger decay essentially implies less track-like events ($\nu_\mu$ contains  more $\nu_3$ than $\nu_e$ does), and the $\theta_{23}$ dependence makes the muon disappearance probability almost flat around $\sin^2\theta_{23} = 0.5$ when compared to the appearance probabilities, this reduction of track-like signatures due to neutrino decay can be compensated
with a lower rate of flavor oscillations in the $\nu_\mu \to \nu_e$ channel, achieved switching the fitted  atmospheric mixing angle from the upper to the lower octant. 

From Fig.~\ref{fig:alpha3sensitivity}, one can read off the expected limits on the invisible neutrino decay from ORCA, summarized in Tab.~\ref{tab:bounds} for 3 and 10 years of running time, too.
Using the relation $\alpha_3=m_3/\tau_3$, it is straightforward to convert the bounds on $\alpha_3$  
 to limits on $\tau_3/m_3$, in order to compare with the  existing limits in the literature, discussed in Sec.~\ref{sec:intro}.  
%.  
 We find that, within the three-neutrino framework, ORCA could improve the current bounds  on  $\alpha_3$ from oscillation experiments by approximately two orders of magnitude.

\begin{table}[!tb]
  \centering
  \renewcommand{\arraystretch}{1.2}
  \begin{tabular}{|c|c|c|}
    \hline
    time & $\alpha_3$ [eV$^2$] & $\tau_3/m_3$ [s/eV]\\
    %\hhline{~--}
    \hline
    3 years & $ < 4.6\times10^{-6}$ &  $> 1.4\times10^{-10}$ \\ \hline
    10 years & $ < 2.6\times10^{-6}$ &  $> 2.5\times10^{-10}$  \\ \hline
  \end{tabular}
  \caption{Expected 90\% C.L. limits from ORCA after 3 and 10 years of running time.}
  \label{tab:bounds}
\end{table}

%%%%%%%%%%%%%%%%%%%%
%%%%%%%%%%%%%%%%%%%%
\subsection{Effect of decay on standard oscillation parameters}

\begin{figure}
\centering
\includegraphics[width=0.85\textwidth]{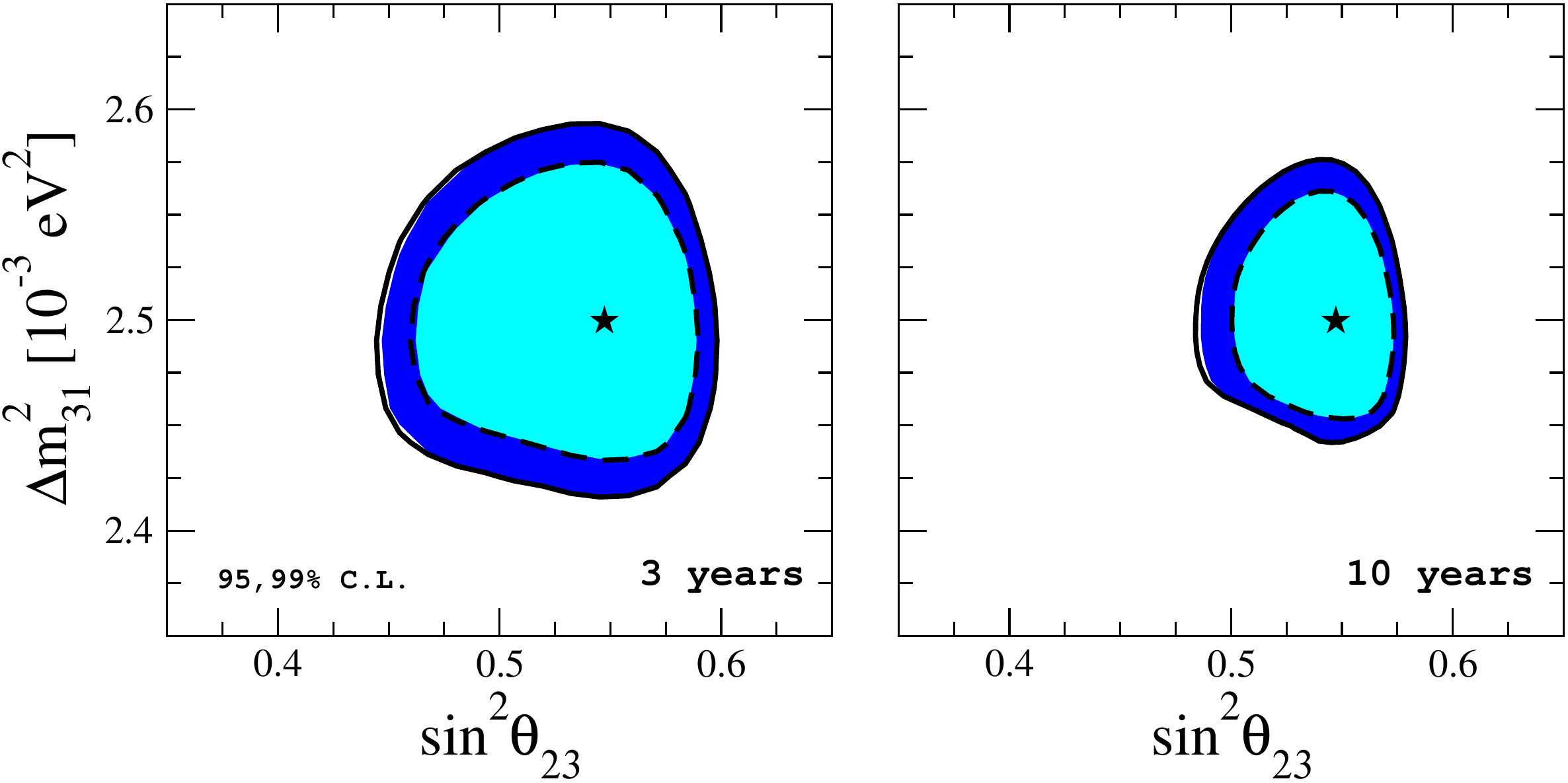}
\caption{
ORCA sensitivity to the atmospheric oscillation parameters in the standard picture (filled regions) and with neutrino decay after marginalizing over $\alpha_3$ (black lines) at 95\% and 99\% C.L. after 3 years (left panel) and 10 years (right panel) of running time.}
\label{fig:standard}
\end{figure}

The existence of invisible neutrino decay could affect the determination of the standard oscillation parameters in ORCA. Here we perform an analysis with the $\chi^2$ function in Eq.~\eqref{eq:chi2}  marginalized  with respect to the decay constant $\alpha_3$,  
\begin{equation}
 \chi^2_{\text{atm}}(\sin^2\theta_{23},\Delta m_{31}^2) = \min_{\alpha_3} \left[ \chi^2 (\sin^2\theta_{23},\Delta m_{31}^2,\alpha_3)\right] \, ,
\end{equation}
for each pair of values $(\sin^2\theta_{23},\Delta m_{31}^2)$, where $\alpha_3$ is varied freely in the fit. The result of our simulation is presented in Fig.~\ref{fig:standard}, where one can see that neutrino decay does not affect the sensitivity of ORCA to the mass splitting, and only worsens very slightly the sensitivity to the atmospheric angle.   

We have also tested the implications of neutrino decay in the ability of ORCA to determine the neutrino mass ordering. 
 To do so, we have simulated data for a given true ordering (TO) of neutrino masses and different assumed true values of the decay constant, $\alpha_3^\text{true}$: 
$N^\text{dat}_{ij}(\sin^2\theta_{23}^\text{BF},\Delta m_{31}^{2\text{ BF}},\alpha_3^\text{true}, \text{TO})$, where we have fixed the oscillation parameters to their best fit value in Tab.~\ref{tab:datasim}.
Then, we  evaluate the $\chi^2$ function in Eq.~\eqref{eq:chi2} assuming the other (wrong) mass ordering (WO) and marginalize over the two oscillation parameters as well as the decay constant $\alpha_3$, with
$N_{ij}(\sin^2\theta_{23},\Delta m_{31}^2,\alpha_3, \text{WO})$.

\begin{figure}
\centering
\includegraphics[width=0.7\textwidth]{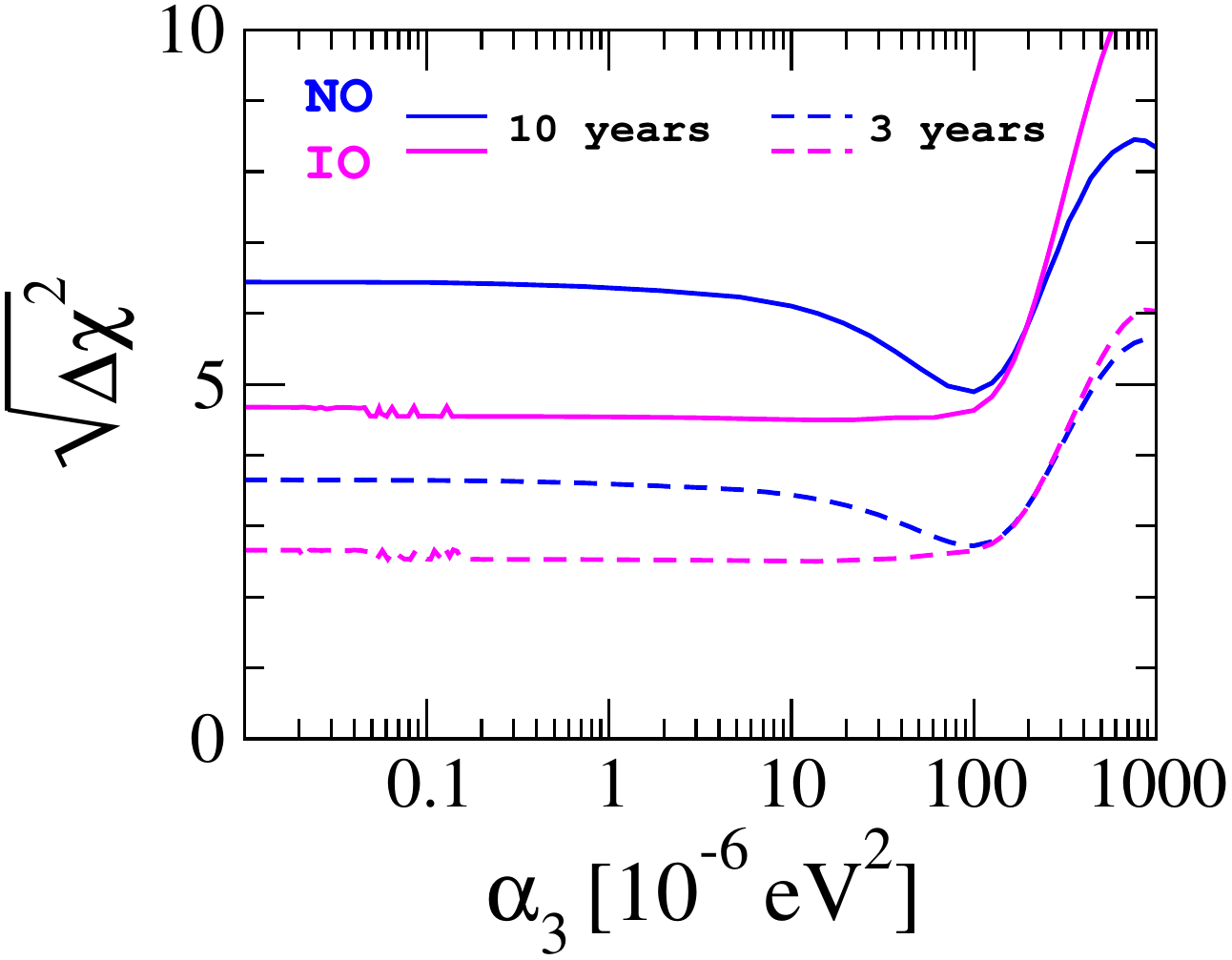}
\caption{
ORCA sensitivity to the neutrino mass ordering as a function of the decay constant $\alpha_3$. Blue (magenta) lines correspond to the case in which the true ordering is assumed to be normal (inverted), and solid (dashed) lines correspond to an exposure time of 10 years (3 years).}
\label{fig:sens-nmo}
\end{figure}

Our results, presented in Fig.~\ref{fig:sens-nmo}, show  ORCA's sensitivity to the ordering of the neutrino mass spectrum when the true ordering is assumed to be normal (blue lines) and inverted (magenta lines). In the limit of very small $\alpha_3$, we recover the standard ORCA sensitivity to the mass ordering, as expected. From the figure, we can  also appreciate a 
 reduction of ORCA's mass ordering discrimination power for $\alpha_3 \sim 10^{-4}\,\mathrm{eV}^2$, when true normal mass ordering is assumed. For values of  
 $\alpha_3$ larger than $10^{-4}\,\mathrm{eV}^2$,  a general increase on the neutrino mass ordering sensitivity is obtained. Note, however, that these values of the decay constant are already excluded by current oscillation limits.
 On the other hand, this enhanced sensitivity to the mass ordering arises from the different oscillatory patterns between NO and IO  induced by the neutrino decay, rather than by matter effects and neutrino flux differences.
In any case, in the region of interest where the experiment is most sensitive, corresponding to  $\alpha_3 \sim 10^{-6}\,\mathrm{eV}^2$, the sensitivity to the neutrino mass ordering is robust and does not show a large dependence on the decay.
Notice that the robustness of ORCA's sensitivity to the neutrino mass ordering is mostly due to the  broad number of baselines and energies  accessible to  atmospheric neutrinos. 
In consequence, this result can not be automatically extended to the case of accelerator-based long-baseline oscillation experiments,  where the fixed baselines and narrow energy  spectrum might hinder the ability of disentangling the effect of flavor oscillations from that of neutrino decay. Dedicated analysis will be required to investigate the impact of neutrino decay on the mass ordering sensitivity of these experiments.

%Although we do not expect this to be a relevant issue for medium and long-baseline oscillation experiments, the effect can be worse for them because of the reduced baselines and narrow detected energies, which hinders the ability of disentangling both effects.

%{\color{red} Notice that the presence of neutrino decay will not affect ORCA's sensitivity to the neutrino mass ordering, at the relevant range of $\alpha_3$ values, thanks to the broad number of baselines and energies that can be accessible when studying atmospheric neutrinos. Although we do not expect this to be a relevant issue for medium and long-baseline oscillation experiments, the effect can be worse for them because of the reduced baselines and narrow detected energies, which hinders the ability of disentangling both effects.
%
%
% In Ref.~\cite{Abrahao:2015rba}, for example, the authors checked this effect for JUNO, finding that the inclusion of neutrino decay {\color{blue} in the fit data barely affects the sensitivity of the experiment to the mass ordering.} {\color{magenta}can only significantly affect the determination of the mass ordering when it is assumed that both, input and fit data, have the presence of decay.}}

\subsection{Other future prospects for invisible neutrino decay}

The sensitivity to the invisible neutrino decay at different future experiments has been recently estimated at several studies. Here we will summarize the most relevant results we have found in the literature.

Ref.~\cite{Choubey:2017dyu} analyzes the sensitivity of the future long-baseline neutrino experiment DUNE. Considering a run of 5 years in neutrino mode plus other 5 years in antineutrino mode,  it is shown that a bound of $\tau_3/m_3 > 4.50\times10^{-11}$ s/eV could be obtained at 90\% C.L. for the case of normal mass ordering.

The authors of Ref.~\cite{Abrahao:2015rba}, on the other hand, study the sensitivity to neutrino decay focusing on reactor experiments. They find that a bound  of $\tau_3/m_3 > 9.1\times10^{-11}$ s/eV at 90\% C.L.\ could be obtained after 5 years of run time in JUNO. Note, however, that in this work the authors performed their analysis marginalizing over 1$\sigma$ ranges of neutrino oscillation parameters and, therefore, a marginalization over the 3$\sigma$ ranges  could worsen this bound.

Concerning atmospheric neutrinos, the sensitivity to the invisible neutrino decay at the proposed India-based Neutrino Observatory (INO) has been analyzed in Ref.~\cite{Choubey:2017eyg}.
After 10 years of data taking, it is found that INO could put a limit of $\tau_3/m_3 > 1.51\times10^{-10}$ s/eV at 90\% C.L.

Therefore, assuming that ORCA will run for ten years, we find that, besides improving the  current bounds on the invisible neutrino decay by approximately two orders of magnitude, it will be, at least, as competitive as DUNE, JUNO and INO, if not a bit better.

%%%%%%%%%%%%%%%%%%%%%%%%%%%%%%%%%%%%%%%%%%%%%%%%%%%%%%%%%%%%%%%%%%%%%%%%%
\section{Conclusions and final remarks}
\label{sec:summary}
%%%%%%%%%%%%%%%%%%%%%%%%%%%%%%%%%%%%%%%%%%%%%%%%%%%%%%%%%%%%%%%%%%%%%%%%%

In this letter we have performed an analysis of the ORCA experiment in the context of invisible neutrino decay. We find that the data on atmospheric neutrinos obtained in ORCA in the GeV energy range could improve the bounds on the decay constant from current oscillation experiments, when including three-neutrino oscillations, by roughly a factor of 100. After ten years, ORCA could constrain the parameter $\tau_3/m_3 > 2.5\times 10^{-10}\text{ s/eV}$ at 90\% C.L. This means that, in the context of future neutrino oscillation experiments, ORCA will better constrain the invisible neutrino decay  in comparison to other experiments such as DUNE, JUNO or INO, although not very significantly.

We also show that the decay of the heaviest neutrino does not affect the determination of the atmospheric oscillation parameters at ORCA.
In particular, the determination of the neutrino mass ordering is very robust against the presence of invisible neutrino decay. 

To conclude this work, we must remark that terrestrial experiments are not the only way to obtain bounds on invisible neutrino decay. 
Complementary constraints on the neutrino lifetime have been derived from astrophysical and cosmological observations which, in many cases, are
significantly more stringent than the bounds presented in this work or from other current or future oscillation experiments. 

Concerning cosmology, the standard Big Bang model predicts the exis\-tence of a relic background of neutrinos that, after the weak decoupling process at MeV temperatures, free stream during later epochs of the expanding universe.
However, if non-standard neutrino interactions exist, depending on their strength, neutrinos could behave like a relativistic fluid, instead of free-streaming particles with anisotropic stress. These interactions could be tested via their imprint on cosmological observables (see e.g.\ 
\cite{Beacom:2004yd,Hannestad:2004qu,Bell:2005dr,Hannestad:2005ex,Basboll:2008fx,Archidiacono:2013dua,Oldengott:2014qra,Forastieri:2015paa}),
such as the anisotropies of the cosmic microwave background. In particular, several cosmological analyses have shown that very stringent bounds
apply to the invisible neutrino decay \cite{Hannestad:2004qu,Hannestad:2005ex,Basboll:2008fx,Archidiacono:2013dua}. For instance, in 
\cite{Archidiacono:2013dua} a cosmological limit on the neutrino lifetime was found at the level of $\log_{10}[(\tau/m)/(\mathrm{s/eV})]\gtrsim 11$,
several orders of magnitude above the values we are considering. However, it is expected that these bounds could be relaxed if only one of the 
neutrino states decays (and the others free stream like in the standard case) or other relativistic non-interacting particles are present, such as thermal
axions.

A future detection of the neutrino mass with cosmological precision data, at the level guaranteed by flavor oscillations, would lead to a huge improvement on the existing limits on the neutrino lifetime by many orders of magnitude. This conclusion applies to the invisible process, and to any kind of neutrino decay, because it is independent of the neutrino decay products \cite{Serpico:2007pt}. Likewise, lifetimes up to a similar level would be tested if the diffuse neutrino flux emitted by all past supernovae is measured in the near future \cite{Ando:2003ie,Fogli:2004gy,Lunardini:2010ab}.

%%%%%%%%% 

\section*{Acknowledgments}

We would like to thank Juande Zornoza for useful discussions about the simulation of the ORCA experiment. 
Work supported by the Spanish grants
FPA2015-68783-REDT, %RENATA
FPA2017-90566-REDC (Red Consolider MultiDark),
FPA2015-65150-C3-1-P, %TARAK
FPA2017-85216-P %AHEP
and SEV-2014-0398 %IFIC
(MI\-NE\-CO/AEI/FEDER, UE), 
as well as PROMETEOII/2014/079 
and PRO\-ME\-TEO/2018/165 (Generalitat Valenciana). 
PFdS and CAT are also supported by the fellowships FPU13/03729 and BES-2015-073593, respectively. 
PFdS acknowledges support by the Vetenskapsr{\aa}det (Swedish Research Council) through contract No. 638-2013-8993 and the Oskar Klein Centre for Cosmoparticle Physics. 
CAT acknowledges the hospitality of the Fermilab Theoretical Physics Department. 
MT acknowledges financial support from MINECO through the Ram\'{o}n y Cajal contract RYC-2013-12438 as well as from the L'Or\'eal-UNESCO \textit{For Women in Science} initiative.

\section*{References}
%\bibliography{bibliography}

\end{document}